# Interest-Related Item Similarity Model Based on Multimodal Data for Top-N Recommendation

Junmei Lv, Bin Song* Member, IEEE, Jie Guo, Member, IEEE, Xiaojiang Du, Senior Member, IEEE, and Mohsen Guizani, Fellow, IEEE

*Abstract*—Nowadays, the recommendation systems are applied in the fields of e-commerce, video websites, social networking sites, etc., which bring great convenience to people's daily lives. The types of the information are diversified and abundant in recommendation systems, therefore the proportion of unstructured multimodal data like text, image and video is increasing. However, due to the representation gap between different modalities, it is intractable to effectively use unstructured multimodal data to improve the efficiency of recommendation systems. In this paper, we propose an end-to-end Multimodal Interest-Related Item Similarity model (Multimodal IRIS) to provide recommendations based on multimodal data source. Specifically, the Multimodal IRIS model consists of three modules, i.e., multimodal feature learning module, the Interest-Related Network (IRN) module and item similarity recommendation module. The multimodal feature learning module adds knowledge sharing unit among different modalities. Then IRN learn the interest relevance between target item and different historical items respectively. At last, the multimodal data feature learning, IRN and item similarity recommendation modules are unified into an integrated system to achieve performance enhancements and to accommodate the addition or absence of different modal data. Extensive experiments on real-world datasets show that, by dealing with the multimodal data which people may pay more attention to when selecting items, the proposed Multimodal IRIS significantly improves accuracy and interpretability on top-N recommendation task over the state-of-the-art methods.

## I. Introduction

WITH the rapid development of the Internet and continuous breakthrough in computing power, online products and contents are exploding [1]-[3], and the formats of online data are various in terms of modalities, which include text, image, etc., providing users with more diversified choices. Meanwhile, overwhelming quantities of products, including clothes, movies, music, news, and books induce the information overload problem. To deal with this problem, recommendation system and search engine emerge. Compared to the search engines, recommendation system is more active to provide users with personalized services and products, creating business benefits.

Mainstream recommendation techniques include Collaborative Filtering (CF), content-based recommendation system and hybrid recommendation system. In particular, CF is based on user preferences for items, reflecting relationships between users or items [4], whereas content-based recommendation system uses the item's own available descriptions to relate them with user preferences. As users pay more and more attention to privacy, most Internet sites cannot obtain detailed user attributes whereas more information about content or products is available, hybrid recommendation system is more versatile and has the advantages of CF and content-based recommendation. CF can learn the commonality of huge group of users and improve the diversity of recommendations. Content-based recommendation can make better use of various content features and have better interpretability. Therefore, the hybrid recommendation method has been adopted in this paper, which combines item-to-item CF (namely item-based CF) and the content-based representation learning to achieve the task of generating new recommendation lists based on users' historical feedback.

In general, each user is associated with a set of historical items which can be tracked by their consumption habits in most online systems. But only items with positive feedback such as clicking, buying, watching can be stored. The early item-based CF research works consider that all historical items of a user contribute equally when modeling user profiles [5]. However, when people selecting new items, items with positive feedback in historical data do not necessarily reflect the equal interest preferences, only some of the interests associated with this item can affect users' behavior [13]. Recently, item similarity

This work was supported by the National Natural Science Foundation of China under Grant (Nos. 61772387 and 61802296), the Fundamental Research Funds for the Central Universities (JB180101), China Postdoctoral Science Foundation Grant (No. 2017M620438), Fundamental Research Funds of Ministry of Education and China Mobile (MCM20170202), and also supported by the ISN State Key Laboratory.

J. Lv, B. Song and J. Guo are with the State Key Laboratory of Integrated Services Networks, Xidian University, 710071, China. B Song is the corresponding author. (e-mail: jmlv_1@stu.xidian.edu.cn, bsong@mail.xidian.edu.cn, jguo@xidian.edu.cn)

X. Du is with Dept. of Computer and Information Sciences, Temple University, Philadelphia PA, 19122, USA (email: dxj@ieee.org)

M. Guizani is with Dept. of Electrical and Computer Engineering, University of Idaho, Moscow, ID 83844, USA (email: mguizani@gmail.com)

models that combine attention mechanisms for top-N recommendation has been developed, for example ACF [13] and NAIS [7]. These models all learn different importance of different historical items for user profile. The difference is that ACF introduces the attention mechanism to indicate user's preference degree in different historical items, NAIS introduces the attention mechanism to differentiate the varying contributions of user's historically interacted for the final prediction. However, the similarity of these methods is that the attention network which is based on the latent factors of users and items, the content information of the item that exists in a large amount is not utilized and the information utilization rate is low.

With the great success of deep learning in Computer Vision (CV) and Natural Language Processing (NLP), research hotspots for CF are mainly focused on neural CF methods [5]-[7]. Using deep neural networks, the state-of-the-art methods have achieved great success compared to traditional CF, like neighbor-based CF and Matrix Factorization (MF). The representation learning of item content is essential for recommendation, including the representation learning of single-modal data [8]-[10] and multimodal data [11], [12]. However, most previous methods simply use a separate neural network model to learn the different modalities' features, and then obtain a common spatial representation only by using simple concatenation or weighted addition, which means the common underlying information between multiple modalities is neglected to some extent.

In this paper, we introduce a novel approach to tackle aforementioned challenges, which is denoted as Multimodal Interest-Related Item Similarity model (Multimodal IRIS). Specifically, we propose a novel multi-layer end-to-end integration framework that simultaneously incorporates visual features and textual features for the task of personalized recommendation on implicit feedback datasets, including a multimodal feature learning module based on sharing knowledge unit, an Interest-Related Network (IRN) optimization module, and an item similarity recommendation module. In order to evaluate the importance of different historical items to characterize users' preference when choosing a new item, we use the multimodal content features of items to learn different interest relevance. As shown in Fig.1, when a user selects a different target movie, his/her preference is reflected in the distribution of interest relevance on the historically viewed movies. We explore how to enable the knowledge sharing on the entire network across modalities and how to construct IRN for predicting based on the multimodal content features of historical items. To the best of our knowledge, such type of recommendation system has not been explored in literature.

In order to evaluate the efficiency of our proposed method, extensive experiments are conducted. Simulation results show that it outperforms previous state-of-the-art methods by NCF to NAIS when applied to the MovieLens dataset[1] and Amazon product dataset [34], which are the most well-studied datasets. In addition, our proposed model can generate more accurate recommendations based on result analyses of the models.

In summary, our contributions are mainly three-fold:

- A Multimodal IRIS framework is proposed to generate recommendations from different modalities. That is, the framework simultaneously learns different modal features and item latent dimensions for the task of personalized recommendation on implicit feedback datasets. By learning the visual information and textual information (such as posters and introductions to movies) which people may pay more attention to when selecting items, the framework will be able to consider the effects of different modalities on item selection at the same time, thus improving the recommendation accuracy and generating recommendations that match user preferences better;
- We propose IRN to learn the interest relevance between target item and different historical items, and add knowledge sharing unit in the multimodal feature learning process, using multimodal features as input to IRN. The multimodal information and IRN improve the interpretability of the recommendation system;
- Our model is constructed into a unified framework, with implementation of multimodal feature learning, IRN optimization, and item similarity recommendation using an end-to-end approach, improving the model scalability over multimodal data.

The rest of the paper is organized as follows. Section II reviews the related works. Section III introduces the proposed Multimodal IRIS model. Section IV provides the evaluation and analysis of the experiments. The final section concludes the whole paper.

## II. RELATED WORKS

In this section, we introduce some works that are related to our researches, including top-N recommendation, representation learning with deep learning and attentional

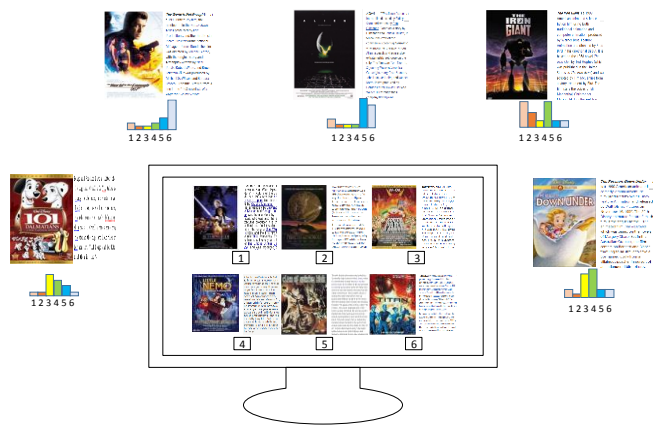

Fig. 1. The illustration of our approach. The movie recommendation is shown as an example. The outer five sets of posters and plots represent different target movies, and the inner six movies constitute the user's historical viewing records. The bar chart is the relevance of the target item above it to the different items in the history. This shows that the target item has different levels of interest in different historical items.

models in recommendation system.

### A. Top-N Recommendation

Early recommendation algorithms mainly focus on CF models, including neighbor-based CF [4] and MF [14], [15]. CF models aim to exploit users' preferences for items (e.g. explicit feedback or implicit feedback) to provide personalized recommendations. Distinct from CF methods based on explicit feedback as a rating prediction task, the works on implicit feedback typically treat CF as a top-N recommendation problem [16], [17].

There is empirical evidence showing that an algorithm optimized for lower rating prediction error does not necessarily result in higher accuracy in terms of top-N recommendation task [18]. Top-N recommendation becomes a more standard task, and has been widely used in E-commerce applications to help the users identify the items that best fit their personal tastes [19]-[23]. The current learning mechanisms for learning the top-N recommendation model from implicit feedback include pointwise and pairwise. Pointwise learning means that each training sample is independently applied to the model for training or testing. In the data set, the label with positive feedback is marked as 1, the label without feedback is marked as 0, and the loss function is the square error of the regression task [5], [24] or the log loss of the classification task [6]. The basic assumption of pairwise methods is that there is a greater interest in the items that have interacted than the item that users do not interact. This method applies both positive and negative feedback samples to the model, where the positive and negative samples are determined by sampling. The objective function is designed to maximize the predicted score intervals between positive and negative samples [7], [25].

### B. Representation Learning with Deep Learning

Deep learning is very effective for extracting high level features from various low-level data, especially unstructured data such as images, text, and audio [26]. In general, a large amount of descriptive information about items and users is available in real-world applications, such as visual descriptions and textual descriptions of clothing. Making full use of multimodal side information provides a way to advance our understanding of items and users, which is conducive to better recommendations. Deep neural networks reduce the efforts in hand-craft feature design and apply to a variety of heterogeneous information such as text, images, audio and even video, and thus has been widely used.

Wang et al. [27] study the problem of enhancing point-of-interest (POI) recommendation with visual contents and propose a visual content enhanced POI recommender system, demonstrating images have potentials to improve the performance of POI recommendation. Chu et al. [28] exploit the effectiveness of visual information (e.g. images of food and furnishings of the restaurant) in restaurant recommendation. The visual features extracted by CNN which can represent restaurant attributes and user preference joint with the text representation are input into common recommendation approaches. Results verify that visual information effectively aids prediction. He et al. [9] design a visual Bayesian personalized ranking (VBPR) algorithm by making use of visual features extracted from product images using (pre-trained) deep networks into matrix factorization. This improves the accuracy of ranking while also avoiding cold start to some extent. A novel Dynamic CF [29] introduces the aesthetic information into a coupled matrix and tensor factorization model for clothing recommendation, in which CNNs are used to learn the images features and aesthetic features. Using gated recurrent units to encode the text sequences into latent factor model, the model in [10] is trained end-to-end on the CF task for multi-task learning including content recommendation and item metadata prediction.

At present, many research works focus on the simultaneous use of multiple modal information for recommendation. Cui et al. [30] infuse visual and textual content of items together to make dynamic predictions, which can well reveal item's characteristics and dynamically capture the user's interest. Chen et al. [8] leverage visual information and textual information in the image to model social media images' semantics and generate personalized image tweet recommendation. Iqbal et al. [31] propose two visually-aware recommendation systems: a visual-only variant and a multimodal variant, and utilize Polylingual LDA to infer style over both modalities. Recommend movies with color histograms of movie posters and still frames [32], showing the visual information provides rich knowledge for understanding movies as well as users' preferences.

### C. Attentional Models in Recommendation System

Attentional models assume that people only need to focus on specific parts of the whole perception space, and can use the model to learn the attention weights distribution of the features with different information quantities. Attention mechanism is typically effective in CV [33] and NLP domains [34], [35], [45]. Moreover, it is also getting more and more applications in recommendation system research.

Neural attention can not only used in conjunction with other deep structures such as MLP, CNN and RNN, but also address some tasks independently [36]. The NIAS [7] applies attention mechanism learning history item's different contributions to modeling user profile when predict the user's preference on an item. He et al. [13] introduce item-level and component-level attention mechanism in CF for multimedia recommendation, named as Attentive CF. We are inspired by the idea that using neural network to learn the attention on different historical items but we adopt different modeling method.

### III. THE PROPOSED MULTIMODAL IRIS

We first briefly review the relevant research that based on item similarity. Then, we introduce the NAIS which is the foundation of this work. Finally, we discuss the proposed Multimodal IRIS in detail.

### A. Standard Item Similarity Method Based on Latent Factor

This method is a hybrid method that learns the latent factor vectors based on neighborhood rules. Given a rating matrix $R$,

whether it is user-item explicit feedback (e.g., ratings, 1-5) or implicit feedback (e.g., purchases/clicks, 0 or 1), $U$ and $I$ are used to denote the sets of users and items, $|U| = m$ and $|I| = n$ are the size of the user set and the item set respectively. Then the prediction score of user $u$ for item $j$ that $u$ has never interacted with is:

$$\hat{r}_{uj} = \frac{\sum_{i \in R_u^+} r_{ui} p_i q_j^T}{|R_u^+|} + b_u + b_j, \quad j \notin R_u^+ \quad (1)$$

where $u$ is the user, $j$ is the target item, $R_u^+$ denotes the set of items that user $u$ has interacted with, $|R_u^+|$ represents the size of $R_u^+$, which is the number of items that user $u$ has interacted with. $b_u$ and $b_j$ represent the biases of a specific user and a specific item respectively. Because explicit feedback is not easy to be collected, implicit feedback systems are more common. For implicit feedback systems, $R$ is a binary matrix, $r_{ui} = 1, i \in R_u^+$. $p_i$ and $q_j$ are the latent factors of $i$ as the historical item and $j$ as the target item, respectively. $p_i^T q_j$ can be considered as an item-item similarity between item $i$ and item $j$.

B. *NAIS (Neural Attentive Item Similarity model)*

NAIS is an improved version of FISM (Factored Item Similarity Model) [5] that uses the neural attention mechanism to learn varying importance of the interacted items, and redesigns the form of attention mechanism to overcome the large variance on the lengths of user interaction histories.

Based on the above Eq.1, FISM adds a hyper-parameter controlling the normalization, the predictive model of FISM is:

$$\hat{r}_{uj} = \left( \frac{1}{|R_u^+|^\alpha} \sum_{i \in R_u^+} p_i \right) q_j^T, \quad j \notin R_u^+ \quad (2)$$

where $\alpha$ is an added hyper-parameter.

Although FISM has achieved good performance, its representation ability can be limited by its equal treatments on all historical items. Therefore, the starting point of NAIS is to model the importance of different historical items in a user profile, which is also a part of our proposed Multimodal IRIS. The predictive model of NAIS is:

$$\hat{r}_{uj} = \left( \frac{1}{|R_u^+|^\alpha} \sum_{i \in R_u^+} \alpha_{ij} p_i \right) q_j^T, \quad j \notin R_u^+ \quad (3)$$

where $\alpha_{ij}$ is an attention factor also importance of item $i$ when predicting the preference of user $u$ for item $j$, learned by a neural network. Compare to FISM, the $\alpha_{ij}$ is varing with different target item $j$, this parameter also considers the importance of different historical items in different target item prediction processes. $\alpha_{ij} = f(p_i, q_j)$, is a function with $p_i$ and $q_j$ as the input. Where $f(\bullet)$ is a Multi-Layer Perception (MLP) and normalized by the softmax function. When normalizing, the author also proposes a smoothing factor to overcome the problem of large differences in the number of items in different user interaction history.

NAIS models different contribution of historical items when predicting different target item by novel use of attention mechanism, and the attention mechanism is redesigned to make it suitable for recommended scenarios, thus providing state-of-the-art performance.

C. *Multimodal IRIS*

We have noticed that NAIS is a framework that is very suitable for extension. We draw on some of its ideas and apply it to unstructured information recommendation scenarios. When the visual information and textual information of the items are available, we explore how to enable the knowledge sharing on the entire network across modalities and how to construct an interest activation network for predicting based on the multimodal content features of history items.

Multimodal data requires feature prefetching prior to end-to-end model learning. The extraction model of visual features can be selected from various Deep CNN models (such as VGG [37], ResNet [38], Inception [39], etc.) pre-trained on the ImageNet dataset to obtain feature vectors, formalized as $v_i = g_v(image_i)$, $g_v(\bullet)$ represents Deep CNN models, $image_i$ represents the original image, $v_i$ represents the extracted visual pre-trained feature vector. The extraction model of textual features can be selected from a variety of NLP models (Doc2Vec [40], BRET [41], etc.) pre-trained on a large number of corpora to obtain feature vectors, formalized as $t_i = g_t(text_i)$, $g_t(\bullet)$ represents NLP models, $text_i$ represents the original text, $t_i$ represents the extracted textual pre-trained feature vector.

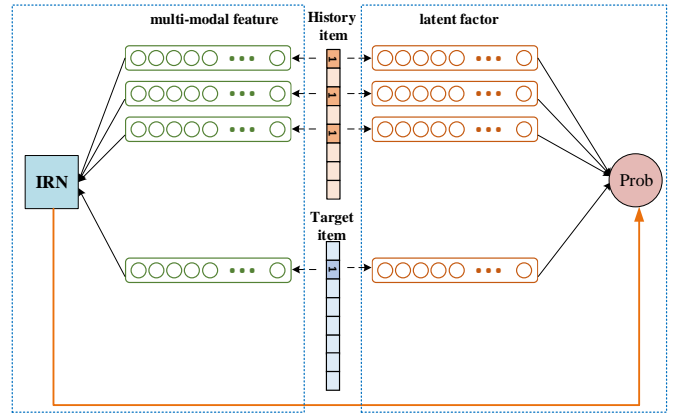

Fig. 2. The structure of our framework. In this framework, the item latent factors are used for similarity calculation and the multimodal features are used to calculate the interest relevance by IRN, so that the preference prediction probability of the user to the target item is integrated.

Before introducing the proposed Multimodal IRIS, we first introduce Image IRIS, which is a special case for Multimodal IRIS dealing with image data. The model uses the visual information in the data to perform effective feature extraction and is used for the input of IRN as shown in Fig.3. We propose IRN for expressing the user's partial interest related to this item based on the relevant historical items activated with the predicted item. Compared to the previous expression of interest values with fixed weights, the IRN will express the changing interest distribution according to the change of the predicted item based on the multimodal content features of the items. Specifically, for Image IRIS, the representation of the IRN is a two-layer perceptron model, and the visual information of the historical items and the target item is taken as input.

$$\alpha_{ij} = \frac{\exp(h^T f(W_1 v_i + W_2 v_j + b))}{\left[\sum_{i \in R_u^+ \setminus \{j\}} \exp(h^T f(W_1 v_i + W_2 v_j + b))\right]^\beta} \quad (4)$$

As mentioned previously, the visual pre-trained feature vector is extracted by the open source pre-trained deep model. Then more refined features, which can be extracted through supervised end-to-end training in Image IRIS, are inputs of the IRN, thereby the visual feature, $V$, is obtained indirectly. $v_i$ and $v_j$ are the corresponding rows, represent the visual feature vector of item $i$ and item $j$ respectively. $\alpha_{ij}$ indicates the interest relevance between the target item $j$ and the historical item $i$. $f(\bullet)$ represents the activation function of the first layer perceptron, we can choose ReLU, sigmoid, etc. Parameter set of perceptron is $\theta = \{W_1, W_2, h, b\}$. Particularly important to note that the hyper-parameter $\beta$ is consistent with NAIS in order to overcome the problem of differences in the number of items in different user histories. The outputs of IRN and the item latent factors is used for similarity recommendation as shown in Fig.2. The formula for generating the prediction is:

$$r_{uj} = \left(\frac{1}{|R_u^+|^\alpha} \sum_{i \in R_u^+ \setminus \{j\}} \alpha_{ij} p_i\right) q_j^T \quad (5)$$

where $P$ and $Q$ are matrices of all item's latent factor vectors, $p_i$ and $q_j$ are the corresponding rows.

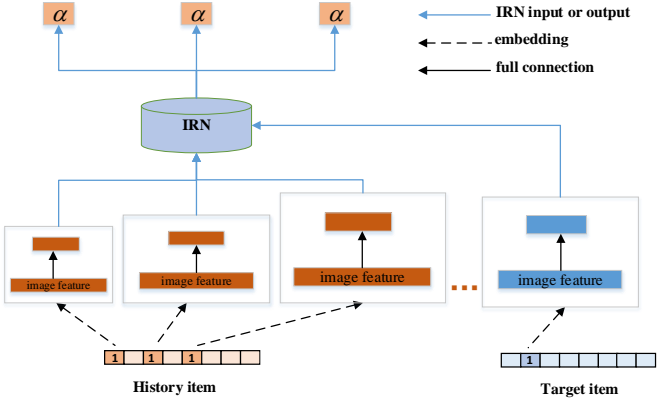

Fig. 3. The interest relevance between items is calculated using image features. Items ID are used as the input, and the output is the interest relevance between the two.

When text data is included in the data source, we first try to make a simple extension based on the Image IRIS, which is weighted addition of the multimodal data at the input of the IRN and recorded as Image-add-Text IRIS. The representation of the IRN is:

$$\alpha_{ij} = \frac{\exp(h^T f(W_1 v_i + W_2 v_j + W_3 t_i + W_4 t_j + b))}{\left[\sum_{i \in R_u^+ \setminus \{j\}} \exp(h^T f(W_1 v_i + W_2 v_j + W_3 t_i + W_4 t_j + b))\right]^\beta} \quad (6)$$

where $t_i$ and $t_j$ represent the textual feature vector of item $i$ and item $j$ respectively. Since the way the above model handles multimodal data consistents with the traditional model, simple weighted addition or splicing loses the knowledge association between different modalities. For this reason, we propose Multimodal IRIS that uses the knowledge sharing unit in feature learning before inputting to IRN as shown in Fig. 4. We assume that hidden layers of supervised visual feature extraction network and supervised textual feature extraction network are connected by a share weight matrix. We enable dual knowledge transfer across modalities by introducing cross connection between one modal network and another, thereby achieving effective integration and synergy of kinds of modality for knowledge enhancement.

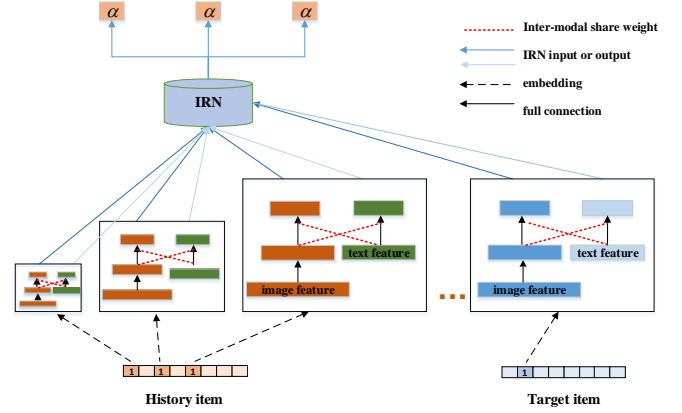

Fig. 4. The interest relevance between items is calculated using multimodal features. Items ID are used as the input, and the output is the interest relevance between the two.

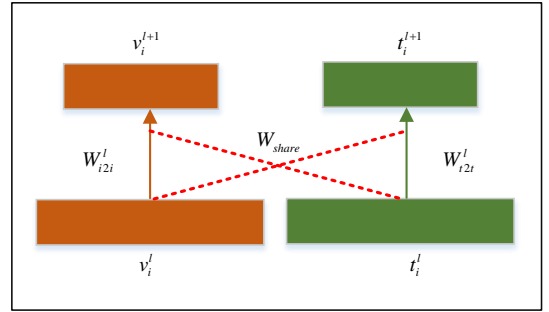

Fig. 5. Knowledge sharing unit between visual feature and textual feature.

$$v_i^{l+1} = \sigma(W_{i2i}^l v_i^l + W_{share} t_i^l) \quad (7)$$

$$t_i^{l+1} = \sigma(W_{t2t}^l t_i^l + W_{share} v_i^l) \quad (8)$$

The above formulas correspond to Fig.5. $v_i^l$, $t_i^l$ represent the visual and textual vector of the previous layer, $W_{share}$ represents the shared hidden layer of between the visual feature and textual feature, designed to capture commonalities between image and text to enhance feature representation, $W_{i2i}^l$ and $W_{t2t}^l$ are parameters required for a typical feedforward network. $v_i^{l+1}$, $t_i^{l+1}$ are the input to IRN.

In the experiment, we further compared the impact of the knowledge sharing unit on the predictive ability of the model.

D. *Model learning*

Due to the greater versatility of the implicit feedback recommendation system, we mainly target Multimodal IRIS optimization based on implicit feedback. For recommendation systems, the parameters of this model are estimated as the minimizer to the following kinds of loss optimization problem:

$$loss_{rmse} = \frac{1}{2} \sum_{u \in U} \sum_{i \in R_u^+} \|r_{ui} - \hat{r}_{ui}\|^2 + \lambda \|\theta\|_F^2 \quad (9)$$

$$loss_{\log} = \frac{1}{2}\sum_{u \in U}\left(\sum_{i \in R_u^+}\log\sigma(\hat{r}_{ui}) + \sum_{i \in R_u^-}\log(1-\sigma(\hat{r}_{ui}))\right) + \lambda\|\theta\|^2 \quad (10)$$

$$loss_{auc} = \frac{1}{2}\sum_{u \in U}\left(\sum_{i \in R_u^+, x \in R_u^-} -\ln(\hat{r}_{ui} - \hat{r}_{ux})\right) + \lambda\|\theta\|^2 \quad (11)$$

where $\hat{r}_{ui}$ is the estimated value for user $u$ to item $i$, $r_{ui}$ is the true feedback for user $u$ on item $i$, $R_u^-$ denotes the set of items that user $u$ has not interacted with. The hyper-parameter $\lambda$ controls the $l_2$ regularization to prevent overfitting, and $\theta$ is a set of trainable parameters. $loss_{rmse}$ and $loss_{\log}$ are to minimize the difference between the original user-item interaction matrix and the reconstructed one in those locations with feedback. $loss_{auc}$ is the BPR pairwise ranking loss [16], $(u, i, x)$ is a sample, where $u$ is a user, $i$ is an item with positive feedback from the user, relatively, $x$ is an item without feedback from the user. The assumption is that interacted user-item pairs should be rated higher than the non-interacted pairs by the model. However, in order to reduce the computational requirements for optimization and keep the balance of train dataset, the negative items are sampled from the set of zero values of $R$.

The $loss_{rmse}$ mentioned above is mainly for the regression loss of the rating prediction task. In addition, we hope to give the predicted probability of each target item in the top-N recommendation list, not just the relative level of the preference possibility. Therefore, we choose the cross-entropy loss for point-wise to optimize the model.

$$\begin{aligned}L = &\frac{1}{2}\sum_{u \in U}\left(\sum_{i \in R_u^+}\log\sigma(\hat{r}_{ui}) + \sum_{i \in R_u^-}\log(1-\sigma(\hat{r}_{ui}))\right) \\ &+ \lambda_1(\|P\|_F^2 + \|Q\|_F^2) \\ &+ \lambda_2(\|W_1\|_F^2 + \|W_2\|_F^2 + \|W_3\|_F^2 + \|W_4\|_F^2) \\ &+ \lambda_3(\|W_{i2i}\|_F^2 + \|W_{t2t}\|_F^2) + \lambda_4\|W_{share}\|_F^2\end{aligned} \quad (12)$$

where $P$ and $Q$ represent two low-rank matrices, where $P \in R^{m \times k}$, $Q \in R^{m \times k}$ and $k << m$. $W_1, W_2, W_3, W_4$ are the parameters of IRN, $W_{share}, W_{i2i}$ and $W_{t2t}$ are the parameters of the feature extraction network, $\lambda_1, \lambda_2, \lambda_3, \lambda_4$ are hyper-parameters used to control the regularization, $\|\bullet\|_F$ is the Frobenius Norm, this norm can be defined as $\|A\|_F = \left\{\sum_{i=1}^m\sum_{j=1}^n|\alpha_{ij}|^2\right\}^{1/2}$, $\alpha_{ij}$ is an element in matrix $A$. During the training, the optimizer can be stochastic gradient descent or its variants such as Adagrad [43], Adam [44] that can adjust the learning rate adaptively. For some scenarios where item data with some modalities missing, if the value of each element in the missing modal data is set to 0, in theory, it can be equivalent to the single modality as input of IRN, and the structure of Multimodal IRIS model does not need to be modified.

## IV. EXPERIMENTS

In this section, we show the comprehensive process of our experiments, including data sets and preprocess, evaluation methodology, comparing baselines, implementation and results analysis. Our algorithm is implemented in Tensorflow-1.9 with Python wrapper and runs at Intel Core i7-6700 and an NVIDIA TITAN X GPU.

### A. Data Sets and Preprocess

We evaluate the performance of the proposed framework on the MovieLens 1M dataset and the subsets of Amazon product dataset namely Clothing & Shoes & Jewelry. Table I shows the statistics of the two datasets.

**MovieLens**: Since the standard MovieLens dataset does not include multimodal data, we use the OMDb API[2] to crawl the corresponding posters and textual plots of the movie as visual information and textual information to validate our proposed Multimodal IRIS. We take users' rating histories as implicit feedback.

**Amazon**: The Amazon dataset contains image information and textual introductions to the product. In order to quickly verify our method, we discard users with less than five interactions and remove items that rating less than 5. We take users' review histories as implicit feedback.

TABLE I
STATISTICS OF EXPERIMENTED DATASETS

| Dataset | User | Item | Interaction | Density |
|---|---|---|---|---|
| MovieLens | 6040 | 3685 | 998034 | 4.48% |
| Amazon | 39387 | 23033 | 278677 | 0.031% |

For each item in the datasets above, we collect one item image and item textual introduction. For image, we extract visual pre-trained features using the ResNet50 model, which is proposed by Kaiming He et al in ImageNet challenge. By using residual modules and regular SGD (which requires reasonable initialization weight), very deep networks can be trained. Models and parameters pre-trained on 1.2 million ImageNet images for this structure are available in the Keras package[3]. In our experiments, we take the output of the flatten layer, to obtain a 2048-dimensional visual feature vector. For text, we extract textual pre-trained features using the BERT. Surprisingly, Google recently introduced the BERT (Bidirectional Encoder Representation from Transformers), which is excellent in many NLP tasks, using Masked LM and Next Sentence Prediction to capture the word and sentence level representation respectively. We use BERT for pre-fetching of textual features, to obtain a 768-dimensional textual feature vector.

### B. Evaluation Methodology

Consistent with the previous research work in recommendation systems [6], [17], we employ the standard leave-one-out protocol. We take all the interaction data as positive samples. For each user, we randomly select one interaction for testing, the rest for training. Then we extract negative samples with equal probability in the non-interactive items, so that the positive and negative sample ratio is 1: K in the training set, where K is a hyper-parameter for controlling sample. In addition, during evaluation, it is impossible to rank all non-interactive items for each user because of great time consumption resulted by the large size of the item set. Therefore, for each positive sample in the test set, 99 negative samples that are not interacted with the user are randomly paired, so that each user in the test set corresponds to 100 interactions, this is following the common approach [6].

---

[2] http://www.omdbapi.com/

[3] https://github.com/keras-team

The typical evaluation metrics on top-N item recommendation based on implicit feedback are Hit Ratio (HR) and Normalized Discounted Cumulative Gain (NDCG). Where HR@N measures whether the positive samples in the test set appear in the top-N recommendation list; NDCG@N considers the position that the positive samples in the test set appear in the top-N recommendation list, where the N is a hyper-parameter. We compare the experimental results under different N. For both metrics, the higher the value is, the better the performance of the model is. The specific form of calculation is:

$$\text{HR @ N} = \frac{\sum_{i=1}^{NUM(user)} co(i)}{NUM(user)} \quad (13)$$

where, $co(i)$ indicates whether the reserved test item is presented on the top-N list of user $i$, including 1 and not including 0, $NUM(user)$ represents the total number of users.

$$\text{NDCG @ N} = \frac{1}{NUM(user)} \sum_{i=1}^{NUM(user)} hits(i)$$

$$hits(i) = \begin{cases} \frac{1}{\log_2(pos_i + 1)} & if \quad co(i) = 1 \\ 0 & if \quad co(i) = 0 \end{cases} \quad (14)$$

where $pos_i$ represents the position of the positive sample in the top-N list of the $i$-th user, $1 \leq pos_i \leq N$.

*C. Baselines*

We compare our proposed Multimodal IRIS with some relevant and state-of-art research methods.

**VBPR** [9]: Visual Bayesian Personalized Ranking incorporates visual signals into predictors of people's opinions. VBPR enhances the performance of matrix decomposition models that only rely on user latent factors and item latent factors. We have modified VBPR to make it suitable for learning using the pointwise log loss function.

**FISM** [5]: This method has been described in detail in the previous section, such as Eq.2. The idea of the fusion latent factor models and neighborhood-based models is consistent with our proposed Multimodal IRIS, so this method is very important as a baseline.

**NCF** [6]: Neural CF leverages an MLP to model the non-linearities between user and item under matrix factorization framework. Due to the powerful representation of neural networks, this model achieves the best performance in the latent factor models.

**NAIS** [7]: Neural Attentive Item Similarity Model designs an attention network for distinguishing different importance of all historical items in a user profile. The item similarity framework using attention mechanism is part of Multimodal IRIS

**Image IRIS**: As in the section III.C, Image IRIS is a special case for Multimodal IRIS with image features as input to the IRN, which can be used to directly compare with VBPR.

**Image-add-Text IRIS**: Image-add-Text IRIS is an extension based on Image IRIS, which is the image and text data weighted addition at the input side of IRN.

**Multimodal IRIS**: It is the complete IRIS framework that uses both image and text features as input to IRN and incorporates a knowledge sharing unit between modalities.

The choice of the above comparison methods makes our experiment more persuasive. VBPR is the most representative latent factor model combined with visual information. NCF is the first to introduce MLP into the latent factor model, breaking through the performance achieved by traditional matrix decomposition techniques. Similarly, in the extraction of various modal deep features, we introduce transformation of features by improved MLP. FISM and NAIS are more direct comparison methods based on the item latent factor, and the application of the attention mechanism in NAIS further achieves the state-of-the-art performance. The comparison between Image-add-Text IRIS and Multimodal IRIS shows the important role of the knowledge sharing unit.

*D. Implementation*

For VBPR, we use the implementation released by Jinhui Tang, The AMR (Adversarial Training towards Robust Multimedia Recommender System) [42] they implemented is based on the prototype of VBPR, so the code of VBPR can be obtained. We have made modifications to the data generation and loss function. For FISM, NAIS and NCF, we use the codes published by the author on GitHub. All methods run on the same TensorFlow platform with Python 3.5. All weight parameters are initialized using Xavier initialization, biases are initialized with 0, and embedding tensors are randomly initialized from Gaussian N (0, 0.01). For optimizing pointwise log loss, the optimizer is Adam [44] with initial learning rate 0.001 with a mini batch size of 512. Positive and negative sampling ratio is consistent with [7], where K is 4. We test the cases of the different embedding size which is 16 and 64. In Multimodal IRIS, the node configuration of hidden layers in the supervised visual feature extraction network is [2048→768→ embedding size], the node configuration of hidden layers in the supervised textual feature extraction network is [768→ embedding size]. The $\alpha$ and $\beta$ in the section III.C have been tested in [5], [7]. We refer to the settings that $\alpha = 0, \beta = 0.8$. We determine the hyper-parameters and whether the training is stopped by means of randomly dividing each user's one interaction into the validation set.

*E. Comparisons with Different Approaches*

In this section, we present experimental results for a variety of methods and discuss the results, the performance difference of various comparison methods can be seen from Table II and Table III. The two tables show that our proposed neural models are better than both the deep models (VBPR and NCF) and the shallow models (FISM and NAIS). With only image information as the input, the proposed Image IRIS consistently outperforms compared with the VBPR on MovieLens and Amazon, these results indicate that the Image IRIS exploits the image information more fully and contributes to the recommendation. NCF improves the recommendation performance by fusing MF and MLP to learn the user–item linear and non-linearities interaction function. The improvement of Multimodal IRIS over NCF confirms the validity of multimodal features and neighbor-based models. Our model achieves a 0.99% improvement in terms of HR@10 and a 1.62% improvement in terms of NDCG@10 compared with NAIS on MovieLens. Meanwhile, on Amazon, the proposed model has a 5.32% improvement in terms of HR@10 and a 4.99% improvement in terms of NDCG@10 compared with NAIS. Therefore, we can conclude that the use of multimodal information can improve the efficiency of recommendation. It is important to point out that the performance of FISM and NAIS differs very little on the Amazon dataset (two orders of magnitude sparser than MovieLens). Due to the extreme sparsity of the data, it is impossible to effectively learn attention networks in NAIS only by latent factors. Compared to MovieLens, the Multimodal IRIS has a more significant effect on Amazon. It reveals the necessity of adding multimodal information when deal with sparser datasets. By comparing Image-add-Text IRIS and Multimodal IRIS, we can see that contrast to weighted addition which is the traditional way of using multiple modal information, the use of knowledge sharing unit in the Multimodal IRIS further increases the evaluation metrics, showing the benefits of knowledge transfer. In summary, our model provides superior performance to existing methods.

Since the FISM, NAIS and the proposed Multimodal IRIS are item similarity methods based on latent factor, we have carried out detailed comparisons under different hyper-parameters on MovieLens. During the experiment, we investigate the impact of the two additional hyper-parameters — the last layer size of the supervised multimodal feature transformation which is the feature embedding size and the length of top-N recommendation list. Table Ⅳ and Table Ⅴ shows the performance of different hyper-parameter setting. We make horizontal comparisons in each table. With the increase of N, the changes of HR and NDCG are in line with the theoretical trend, and both metrics have achieved some improvements. The more active the user is, the longer the list of acceptable recommendations is, and the more accurate the recommendation is. We perform a vertical comparison of the same metric in each table. With the same settings, the metric values show an upward trend from top to bottom, showing that the performance of this group of methods is gradually improved. By comparing the two tables, it is shown that the performance of every model above is significantly improved when the feature embedding size rising from 16 to 64. This is a reasonable result because the increase in embedding size enhances the ability to features expression. In theory, the same method, the stronger the feature representation ability is, the higher the HR and NDCG should be. In addition, because the work of multimodal feature pre-extraction can be done offline, compared to NAIS, the increased time complexity of Multimodal IRIS in single training and prediction is only due to the shallow feedforward network for supervised feature extraction. But as shown in Fig. 6, the training loss of Multimodal IRIS decreases fastest under the same number of iterations. In our experiments, our model training process required fewer iterations. So overall, under the current computing resources, the increase in time complexity is not obvious.

TABLE Ⅱ
COMPARISON RESULTS OF DIFFERENT METHODS ON MOVIELENS

| Method | HR@10 | NDCG@10 |
|---|---|---|
| VBPR | 0.7852 | 0.5386 |
| FISM | 0.8136 | 0.5750 |
| NCF | 0.8101 | 0.5519 |
| NAIS | 0.8194 | 0.5736 |
| Image IRIS | 0.8247 | 0.5876 |
| Image-add-Text IRIS | 0.8252 | 0.5875 |
| Multimodal IRIS | **0.8293** | **0.5898** |

TABLE Ⅲ
COMPARISON RESULTS OF DIFFERENT METHODS ON AMAZON

| Method | HR@10 | NDCG@10 |
|---|---|---|
| VBPR | 0.3571 | 0.2017 |
| FISM | 0.4080 | 0.2592 |
| NCF | 0.3620 | 0.2095 |
| NAIS | 0.4099 | 0.2425 |
| Image IRIS | 0.4203 | 0.2573 |
| Image-add-Text IRIS | 0.4530 | 0.2884 |
| Multimodal IRIS | **0.4631** | **0.2924** |

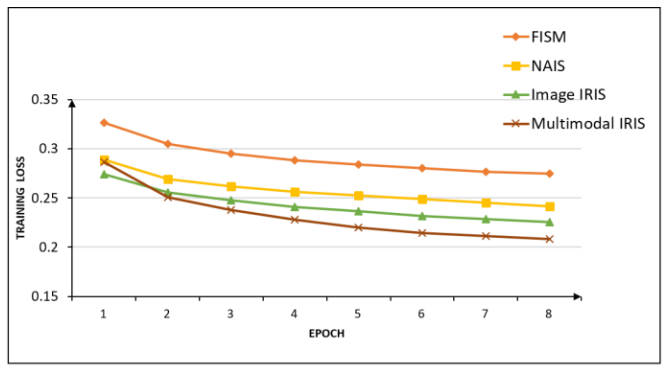

Fig. 6. The training loss of various methods varies with the number of iterations.

TABLE IV
COMPARISON RESULTS OF VARIOUS ITEM SIMILAR MODELS BASED LATENT FACTOR

| Methods | Embedding size:16 | | | |
|---|---|---|---|---|
| | HR@10 | NDCG@10 | HR@20 | NDCG@20 |
| FISM | 0.7914 | 0.5384 | 0.8945 | 0.5597 |
| NAIS | 0.7992 | 0.5523 | 0.9025 | 0.5760 |
| Image IRIS | 0.8083 | 0.5556 | 0.9040 | 0.5776 |
| Multimodal IRIS | **0.8113** | **0.5639** | **0.9053** | **0.5798** |

TABLE V
COMPARISON RESULTS OF VARIOUS ITEM SIMILAR MODELS BASED LATENT FACTOR

| Methods | Embedding size:64 | | | |
|---|---|---|---|---|
| | HR@10 | NDCG@10 | HR@20 | NDCG@20 |
| FISM | 0.8136 | 0.5750 | 0.9066 | 0.5941 |
| NAIS | 0.8194 | 0.5736 | 0.9109 | 0.6025 |
| Image IRIS | 0.8247 | 0.5876 | 0.9147 | 0.6113 |
| Multimodal IRIS | **0.8293** | **0.5898** | **0.9149** | **0.6123** |

It is worth noting that the performance of the Multimodal IRIS predictive model differs when using different text preprocessing models. Due to the amazing performance of BERT on other various NLP tasks recently, we specifically investigated the performance differences caused by the use of BERT and Doc2Vec as text preprocessing models. As can be seen from the Table VI, BERT has slightly improved the performance of Multimodal IRIS. Doc2Vec, also called documents embedding, is an unsupervised algorithm that can get the vector expression of documents, which is an extension of word2vec. The learned vectors can be used to find the similarity between the documents by calculating the distance. Textual feature extraction can be obtained by using the Doc2Vec tool pre-trained document vector in the gensim package [4]. Compared to Doc2vec using Paragraph Vector framework to predict the next word in a context, BERT takes two tricks, one is Masked LM which refers to mask some percent of words from the input then reconstruct those words from context, the other is Next Sentence Prediction which is to concatenate two sentences A and B and predict whether B actual comes after A in the original text, learning the semantic relationship between the words or the sentences. Thereby more fully exploiting the potential interest relevance in the text and achieving better performance in the recommendation system.

TABLE VI
COMPARISON OF FINAL RESULTS USING DIFFERENT TEXT PREPROCESSING MODELS ON MOVIELENS

| Methods | Embedding size:16 | | Embedding size:64 | |
|---|---|---|---|---|
| | HR@10 | NDCG@10 | HR@10 | NDCG@10 |
| BERT | **0.8113** | **0.5639** | **0.8293** | **0.5898** |
| Doc2Vec | 0.8088 | 0.5573 | 0.8270 | 0.5891 |

---

[4] https://radimrehurek.com/gensim/

TABLE VII
A LIST OF RECOMMENDATIONS FOR THE USER WITH ID 10 AND HISTORICAL ITEMS OF HIGH INTEREST RELEVANCE

| User ID:10 | |
|---|---|
| **Top-3 Recommendation** | **The three historical items with the highest interest relevance** |
| 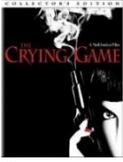 **Crying game (0.857)** | 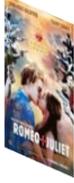 Romeo+juliet (0.679)    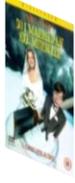 So I Married an Axe Murderer (0.584)    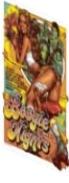 Boogie Night (0.270) |
| 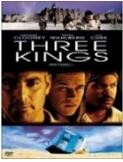 **Three kings (0.830)** | 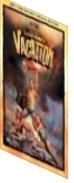 Vacation (0.705)    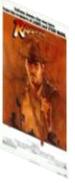 Raiders of the Lost Ark (0.343)    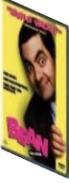 Bean (0.254) |
| 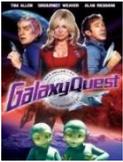 **GalaxyQuest (0.767)** | 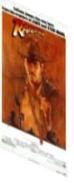 Raiders of the Lost Ark (0.819)    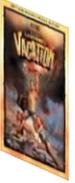 Vacation (0.635)    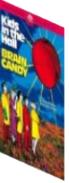 Kids in the Hall:Brain Candy (0.472) |

TABLE VIII
A LIST OF RECOMMENDATIONS FOR THE USER WITH ID 3298 AND HISTORICAL ITEMS OF HIGH INTEREST RELEVANCE

| User ID: 3298 | | | |
|---|---|---|---|
| **Top-3 Recommendation** | **The three historical items with the highest interest relevance** | | |
| 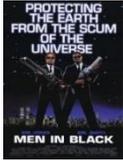 **Man in Black (0.760)** | 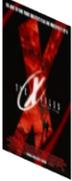 The X-Files (0.655) | 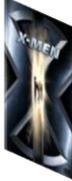 X-MEN (0.559) | 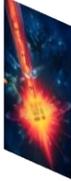 Star Trek (0.484) |
| 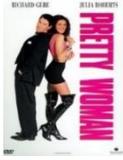 **Pretty Woman (0.694)** | 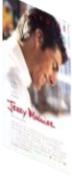 Jerry Maguire (0.771) | 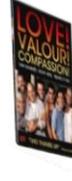 Love!Valour!Compassion! (0.498) | 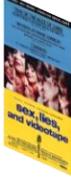 Sex,Lies,andVideotape (0.488) |
| 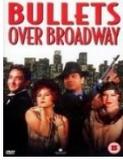 **Bullets over broadway (0.675)** | 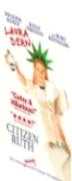 Citizen Ruth (0.651) | 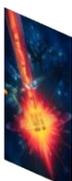 Star Trek (0.483) | 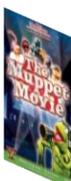 The Muppet Movie (0.388) |

TABLE IX
A LIST OF RECOMMENDATIONS FOR THE USER WITH ID 3424 AND HISTORICAL ITEMS OF HIGH INTEREST RELEVANCE

| User ID:3424 | | | |
|---|---|---|---|
| **Top-3 Recommendation** | **The three historical items with the highest interest relevance** | | |
| 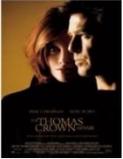 **The Thomas Crown Affair (0.746)** | 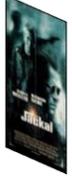 The Jackal (0.589) | 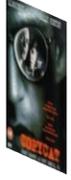 Copycat (0.525) | 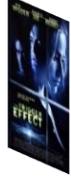 The Trigger Effect (0.470) |
| 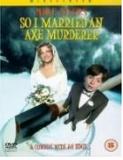 **So I Married an Axe Murderer (0.642)** | 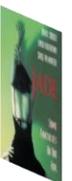 Jade (0.700) | 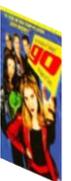 Go (0.632) | 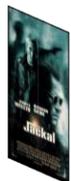 The Jackal (0.603) |
| 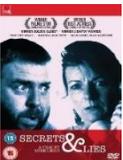 **Secrets & Lies (0.575)** | 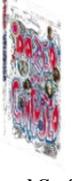 Dazed and Confused (0.868) | 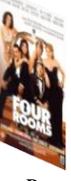 Four Rooms (0.443) | 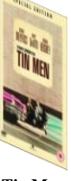 Tin Men (0.438) |

We sort the recommended item lists based on predicted probability and interest relevance. Due to space constraints, we randomly extract three users whose ID are 10 (as shown in Table Ⅶ), 3298 (as shown in Table Ⅷ), and 3424 (as shown in Table Ⅸ) in MovieLens, the top three items of their top-10 recommendation lists and the top-3 historical items with the highest interest relevance for each target item are displayed. In this experiment, item refers to movie. The first column is the recommended movies, each showing poster, name, and predicted probability. The three movies in the second column of each recommended movie row are the most relevant movies in the user historical viewing records computed by the IRN, each of which displays the poster, name, and interest relevance. Table Ⅶ, Table Ⅷ and Table Ⅸ show that the recommendation lists are diverse based on historical viewing records, and the same historical item has different interest relevance to different target items. The multimodal information and interest relevance improve the interpretability of the recommendation system.

## V. Convulsion

In this paper, we propose a novel framework termed as Multimodal IRIS by introducing a knowledge sharing unit and an IRN to an item similarity model for the top-N recommendation. By learning the visual information and textual information, the framework will be able to consider the effects of different modalities at the same time. The knowledge sharing unit has the ability to achieve knowledge transfer among multimodal features and further enhance the performance of the model. IRN with multimodal features as input models the interest relevance between different historical items and target item, not only helps improve recommendation accuracy but also improves interpretability of the recommendation results. Comparing with FISM, NAIS, Image IRIS, etc., the experimental results on the MovieLens and Amazon datasets demonstrate that the proposed Multimodal IRIS model can give more accurate top-N recommendation lists. The recommendation results from some randomly selected users show that the lists that obtained by the model achieve diversity. Also, the addition of multimodal information and the different interest relevance of historical items further enhance the interpretability of the recommendation results. As part of future work, we are interested in investigating how to eliminate the semantic gap between structured information and unstructured information in recommendation system.

BIOGRAPHY

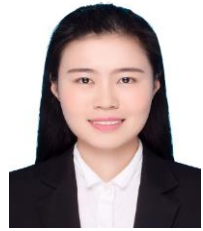

**Junmei Lv** received her B.S. degree in communication engineering from the Yanshan University, Qinhuangdao, China in 2016. She is currently working towards her M.S. degree with electronics and communication engineering from Xidian University, Xi'an, China. Her research interests include machine learning, recommendation system, and data mining.

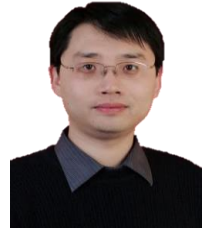

**Bin Song** received his BS, MS, and PhD in communication and information systems from Xidian University, Xi'an, China in 1996, 1999, and 2002, respectively. In 2002, he joined the School of Telecommunications Engineering at Xidian University where he is currently a professor of communications and information systems. He is also the associate director at the State Key Laboratory of Integrated Services Networks. He has authored over 50 journal papers or conference papers and 30 patents. His research interests and areas of publication include video compression and transmission technologies, video transcoding, error- and packet-loss-resilient video coding, distributed video coding, and video signal processing based on compressed sensing, big data, and multimedia communications.

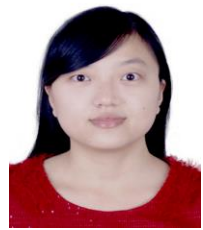

**Jie Guo** received the B.S. degree in communication engineering from Zhengzhou University, Zhengzhou, China, in 2011, and the Ph.D. degree in communication and information systems from Xidian University, Xi'an, China, in 2017. From 2015 to 2016, she was a visiting student with Carleton University, Canada. She is currently a Post-Doctoral Researcher with Xidian University. Her research interests include video compression, transmission, multimodal data fusion, and compressed video sensing.

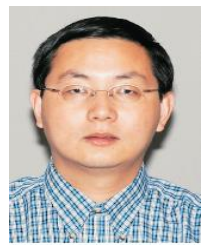

**Xiaojiang (James) Du** is a tenured professor in the Department of Computer and Information Sciences at Temple University, Philadelphia, USA. Dr. Du received his M.S. and Ph.D. degrees in electrical engineering from the University of Maryland College Park in 2002 and 2003, respectively. His research interests are wireless communications, wireless networks, security, and systems. He has authored over 300 journal and conference papers in these areas as well as a book, published by Springer. He won the best paper award at IEEE GLOBECOM 2014 and the best poster runner-up award at the ACM MobiHoc 2014. Dr.

Du served as the lead Chair of the Communication and Information Security Symposium of the IEEE International Communication Conference (ICC) 2015 and a Co-Chair of Mobile and Wireless Networks Track of IEEE Wireless Communications and Networking Conference (WCNC) 2015. He is (was) a Technical Program Committee (TPC) member of several premier ACM/IEEE conferences. Dr. Du is a Senior Member of IEEE and a Life Member of ACM.

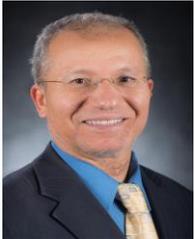

**Mohsen Guizani** (S'85–M'89–SM'99–F'09) received his bachelor's (with distinction) and master's degrees in electrical engineering and master's and doctorate degrees in computer engineering from Syracuse University, Syracuse, NY, USA in 1984, 1986, 1987, and 1990, respectively. He is currently a professor and the ECE Department chair at the University of Idaho. Previously, he served as the associate vice president of Graduate Studies, Qatar University, chair of the Computer Science Department, Western Michigan University, and chair of the Computer Science Department, University of West Florida. He also served in academic positions at the University of Missouri-Kansas City, University of Colorado-Boulder, Syracuse University, and Kuwait University. His research interests include wireless communications and mobile computing, computer networks, mobile cloud computing, security, and smart grid. He currently serves on the editorial boards of several international technical journals and is the founder and the editor-in-chief of the Wireless Communications and Mobile Computing journal (Wiley). He is the author of nine books and more than 400 publications in refereed journals and conferences. He guest-edited several special issues in IEEE journals and magazines. He also served as a member, chair, and general chair at several international conferences. He was selected as the Best Teaching Assistant for two consecutive years at Syracuse University. He received the Best Research Award from three institutions. He was the chair of the IEEE Communications Society Wireless Technical Committee and the chair of the TAOS Technical Committee. He served as the IEEE Computer Society Distinguished Speaker from 2003 to 2005.